\documentstyle[12pt]{article}

\def\a{\alpha}
\def\b{\beta}
\def\c{\chi}
\def\d{\delta}
\def\e{\epsilon}

\def\f{\phi}
\def\vf{\varphi}
\def\g{\gamma}
\def\h{\eta}
\def\j{\psi}

\def\l{\lambda}
\def\m{\mu}
\def\n{\nu}

\def\q{\theta}
\def\r{\rho}
\def\s{\sigma}
\def\t{\tau}

\def\x{\xi}

\def\F{\Phi}
\def\G{\Gamma}
\def\J{\Psi}

\def\inbar{\vrule height1.5ex width.4pt depth0pt}
\def\rlx{\relax\leavevmode}
\def\I{\leavevmode\hbox{\small1\kern-3.8pt\normalsize1}}
\def\openone{\leavevmode\hbox{\small1\kern-3.3pt\normalsize1}}
\def\Ione{\rlx{\rm 1\kern-2.7pt l}}
\font\cmss=cmss10
\font\cmsss=cmss10 at 7pt
\def\ZZ{\rlx\leavevmode
             \ifmmode\mathchoice
                    {\hbox{\cmss Z\kern-.4em Z}}
                    {\hbox{\cmss Z\kern-.4em Z}}
                    {\lower.9pt\hbox{\cmsss Z\kern-.36em Z}}
                    {\lower1.2pt\hbox{\cmsss Z\kern-.36em Z}}
               \else{\cmss Z\kern-.4em Z}\fi}
\def\Ik{\rlx{\rm I\kern-.18em k}}  % Yes, I know. This ain't capital.
\def\IC{\rlx\leavevmode
             \ifmmode\mathchoice
                    {\hbox{\kern.33em\inbar\kern-.3em{\rm C}}}
                    {\hbox{\kern.33em\inbar\kern-.3em{\rm C}}}
                    {\hbox{\kern.28em\sinbar\kern-.25em{\rm C}}}
                    {\hbox{\kern.25em\ssinbar\kern-.22em{\rm C}}}
             \else{\hbox{\kern.3em\inbar\kern-.3em{\rm C}}}\fi}
\def\IP{\rlx{\rm I\kern-.18em P}}
\def\IR{\rlx{\rm I\kern-.18em R}}
\def\IN{\rlx{\rm I\kern-.20em N}}

\def\llsymbol#1{\@llsymbol{\@nameuse{c@#1}}}
\def\@llsymbol#1{\ifcase#1\or {}\or {'}\or {''}\or {'''}\or
   {''''}\or {'''''}\or  \else\@ctrerr\fi\relax}

\newcounter{contador}

\newcommand{\ol}\overline
\newcommand{\ti}\tilde
\newcommand{\wt}\widetilde
\newcommand{\wh}\widehat
\newcommand{\bv}\breve
\newcommand{\dg}\dagger

\newcommand{\sqed}{\mbox{\scriptsize SQED}}
\newcommand{\sd}{\mbox{\scriptsize SD}}
\newcommand{\C}{^{\mbox{\scriptsize c}}}
\newcommand{\sC}{\mbox{\scriptsize c}}

\newcommand{\QED}{QED$_{\mbox{\scriptsize 2+2}}$}

\newcommand{\DDdd}{$D$$=$$2$$+$$2$}
\newcommand{\Ddd}{$D$$=$$1$$+$$2$}

\newcommand{\aand}{\;\;\;\mbox{and}\;\;\;}

\newcommand{\be}{\begin{equation}}
\newcommand{\ee}{\end{equation}}
\newcommand{\bl}{\begin{eqnarray}&}
\newcommand{\el}{&\end{eqnarray}}
\newcommand{\bq}{\begin{eqnarray}}
\newcommand{\eq}{\end{eqnarray}}

\newcommand{\sx}{\sigma_x}
\newcommand{\sy}{\sigma_y}
\newcommand{\sz}{\sigma_z}
\newcommand{\sm}{{\s}^{\m}}

\newcommand{\ad}{{\dot\alpha}}
\newcommand{\bd}{{\dot\beta}}

\newcommand{\gm}{{\gamma}^m}

\newcommand{\uptad}{\widetilde\theta^{\dot\alpha}}

\newcommand{\qt}{\tilde\theta}
\newcommand{\qwt}{\widetilde\theta}
\newcommand{\ov}{\overline}
\newcommand{\pa}{\partial}

\def\sl#1{\rlap{\hbox{$\mskip 1 mu /$}}#1}	% good slash for lower case
\def\ssl#1{\rlap{\hbox{$ {\scriptstyle /}$}}#1}

\begin{document}

\title{\Large \bf $N\,$=1 super-Chern-Simons coupled to parity-preserving
matter from Atiyah-Ward space-time}

\author{{\it M. A. De Andrade}{\thanks{On leave of absence from the Brazilian
Centre for Research in Physics (CBPF), Department of Field Theory and Particles
(DCP), e-mail: marco@cbpfsu1.cat.cbpf.br .}}~,~ {\it O. M. Del Cima}{\thanks{On
leave of absence from the Brazilian Centre for Research in Physics (CBPF),
Department of Field Theory and Particles (DCP), e-mail:
oswaldo@cbpfsu1.cat.cbpf.br .}}~ and~ {\it L. P. Colatto}{\thanks{e-mail:
colatto@ictp.trieste.it}} \\
{\normalsize International Centre for Theoretical Physics (ICTP)} \\
{\normalsize High Energy Section (HE)}\\
{\normalsize Strada Costiera 11, 34100} \\
{\normalsize Trieste, Italy.}}

\date{}

\maketitle

\begin{abstract}
In this letter, we present the Parkes-Siegel formulation for the massive
Abelian $N$$=$$1$ super-{\QED} coupled to a self-dual supermultiplet, by
introducing a chiral multiplier superfield. We show that after carrying out a
suitable dimensional reduction from ($2$$+$$2$) to ($1$$+$$2$) dimensions, and
performing some necessary truncations, the simple supersymmetric extension of
the
${\tau}_{3}$QED$_{1+2}$ coupled to a Chern-Simons term naturally comes out.
\end{abstract}

The issue of self-duality has deserved a great deal of attention since a
self-dual Yang-Mills theory in Atiyah-Ward space-time ($2$$+$$2$ dimensions)
{\cite{selfdym}} has been pointed out as a source for
various integrable models in lower dimensions, according to a conjecture by
Atiyah and Ward {\cite{award}}.

Recently, the simple supersymmetric version of the self-dual Yang-Mills theory
and  self-dual supergravity model in Atiyah-Ward space-time has been formulated
by Gates, Ketov and Nishino {\cite{gatesketnish2}}. Also, by a suitable
dimensional reduction proposed by Nishino, $N$$=$$1$ and $N$$=$$2$
super-Chern-Simons theories in {\Ddd} were generated from $N$$=$$1$ and
$N$$=$$2$ super-self-dual Yang-Mills theories in {\DDdd} {\cite{drscs}}.

In the last years, 3-dimensional field theories {\cite{djt}} have been
well-motivated in view of the possibilities of providing
a gauge-theoretical foundation for the description of condensed matter
phenomena, such as high-$T_{c}$ superconductivity {\cite{domavro}},
where the QED$_{3}$ and ${\tau}_{3}$QED$_{1+2}$ {\cite{domavro,qedtau3}}
are some of the theoretical approaches used to understand more deeply about
high-$T_{c}$ materials. The finiteness on the Landau gauge of Chern-Simons
theories {\cite{finiteness}} is also an interesting result that motivates the
study of 3-dimensional gauge
theories.

The relationship between massive
Abelian $N$$=$$1$ super-{\QED} in Atiyah-Ward space-time and $N$$=$$1$
super-${\tau}_{3}$QED in {\Ddd} has already been investigated by carrying out a
dimensional reduction {\it{\`a la}} Scherk from ($2$$+$$2$) to ($1$$+$$2$)
dimensions and by performing some suitable supersymmetry-preserving truncations
{\cite{stau3qed,trab1}}.

The purpose of this letter is to show that $N$$=$$1$ super-${\tau}_{3}$QED
coupled to a super-Chern-Simons term in {\Ddd} can be generated from the
massive Abelian $N$$=$$1$ super-{\QED} {\cite{stau3qed,trab1}} coupled to a
self-dual supermultiplet by using the Parkes-Siegel formulation in {\DDdd}
{\cite{parkessiegel}}. The dimensional reduction used here to show the
relationship between the models previously mentioned was proposed by Nishino in
Ref.{\cite{drscs}}. Also, some suitable supersymmetry-preserving truncations
are needed in order to suppress non-physical propagating modes as well as to
keep a simple supersymmetry in {\Ddd}.

To introduce mass to the matter sector in {\DDdd}, without breaking
gauge-symmetry, we have to deal with four scalar superfields: a pair of chiral
and a pair of anti-chiral superfields; the members of each pair have opposite
$U(1)$-charges {\cite{stau3qed,trab1}}.
The Parkes-Siegel formulation for the massive Abelian $N$$=$$1$ super-{\QED}
coupled to a self-dual supermultiplet, by introducing a chiral multiplier
superfield, is described by the action :
\footnote{We are adopting in this letter, $\eta_{\m \n}$$=$$(+,-,-,+)$, for the
A-W space-time metric, $ds$$\equiv$$d^4xd^2\q$, $d\wt{s}$$\equiv$$d^4xd^2\qwt$
and $dv$$\equiv$$d^4xd^2{\q}d^2\qwt$, where $\q$ and $\qwt$ are Majorana-Weyl
spinors. Also, the supersymmetry covariant derivatives are defined by :
$D_{\a}$$=$$\pa_{\a}$$-$$i\sl\pa_{\a \ad}\uptad$ and
${\wt{D}_{\ad}}$$=$$\wt{\pa}_{\ad}$$-$$i\sl\wt{\pa}_{\ad \a}\q^{\a}$. For
more details about notational conventions in {\DDdd} and {\Ddd}, see
ref.{\cite{stau3qed,trab1}}.}
\bq
S_{\sqed}^{\sd}\!\!\!&=&\!\!\!-\int{ds}\;\Xi^{\sC} W +
\int{dv}\;\left(\J^{\dg}_+e^{4qV}{\wt X}_+ + \J^{\dg}_-e^{-4qV}{\wt X}
_-\right)+\nonumber\\&& +
\,i\,m\left(\int{ds}\; \J_+\J_--\int{d\wt{s}}\;{\wt X}_+{\wt X}_-\right) +
\mbox{h.c.} \;\;\;\;\;, \label{masssdqed}
\eq
where $q$ is a real dimensionless coupling constant and $m$ is a real parameter
with
dimension of mass. The $+$ and $-$ subscripts in the matter superfields refer
to their respective $U(1)$-charges.

In the action (\ref{masssdqed}), the chiral ($\J_\pm$), the anti-chiral (${\wt
X}_\pm$) and the chiral multiplier ($\Xi$) superfields are defined as follows:
\be
\J_{\pm}(x,\q,\wt{\q})=e^{i\qt\ssl{\tilde\pa}\q}
\left[A_{\pm}(x)+i\q\j_{\pm}(x)+i\q^2F_{\pm}(x) \right]
\;\;\;,\;\;\;\;\wt{D}_\ad \J_{\pm}=0 \;\;\;\;, \label{psi+}
\ee
\be
{\wt X}_{\pm}(x,\q,\wt{\q})=e^{i\q\ssl{\pa}\qt}
\left[B_{\pm}(x)+i\qwt\wt\c_{\pm}(x)+i\qwt^{2}G_{\pm}(x) \right]
\;\;\;,\;\;\;\;D_\a \wt X_{\pm}=0 \;\;\;\;, \label{chi+}
\ee
\be
\Xi_{\a}(x,\q,\wt{\q})=e^{i\qt\ssl{\tilde\pa}\q}\left[A_{\a}(x)+
{\q^\b}\left(\e_{\a\b}{E}(x)-
\s^{\m\n}_{\a\b}H_{\m\n}(x)\right)+
i\q^2{F}_{\a}(x)\right]\;\;\;,\;\;\;\;\wt{D}_\bd \Xi_\a=0 \;\;\;\;,
\label{superspinor}
\ee
where, $A_\pm$ and $B_\pm$ are complex scalars, $\j_\pm$ and $\wt{\c}_\pm$ are
Weyl
spinors, $F_\pm$ and $G_\pm$ are complex auxiliary scalars, $A_{\a}$ is a Weyl
spinor, $E$ is a complex scalar, $H_{\m\n}$ is a
complex antisymmetric rank-2 tensor and ${F}_{\a}$ is a Weyl auxiliary spinor.

In the Wess-Zumino gauge {\cite{wesszugauge}}, a complex {\it vector}
superfield, $V$, is written as
\be
V(x,\q,\qwt)=\frac12i\q\s^\m\qwt
B_\m(x)-\frac12\qwt^2\q\l(x)-\frac12\q^2\qwt\wt\r(x)-\frac14\q^2\qwt^2
D(x)\;\;\;\;, \label{supervector}
\ee
where $D$ is a complex auxiliary scalar, $\l$ and $\wt\r$ are Weyl spinors and
$B_\m$ is a {\it complex} vector field.

The field-strength superfields, $W_\a$ and ${\wt W}_\ad$, defined by
\be
W_\a=\frac12{\wt D}^2D_\a V \aand
\wt{W}_\ad=\frac12{D}^2\wt{D}_\ad V \;\;,
\ee
respectively, satisfy the chiral and anti-chiral conditions,
$\wt{D}_{\bd}W_\a$$=$$0$ and $D_{\b}{\wt W}_\ad$$=$$0$ .

By adopting the Wess-Zumino gauge and considering the superfields defined
above, the following component-field
action stems from the superspace action of eq.(\ref{masssdqed}) :
\bq
S_{\sqed}^{\sd}\!\!\!&=&\!\!\!\int{d^4x}\left\{-\frac12
H^{*}_{\m\n}\left(G^{\m\n}-\frac12 \epsilon^{\m\n\r\s}G_{\r\s} \right)
-i \biggl({A\C}{\sl\pa}\wt\r+F\C\l \biggr)- E^*D \; +\right.
\nonumber\\
&&
-F_+^{*}G_+ - A_+^{*}\Box B_+ - {1\over2} i {\j _+\C} {\sl{\pa}} \wt{\c}_+ -
qB_{\m} \left({1\over2} i {\j_+ \C}   \sm \wt{\c}_+ + A_+^*{\pa}^{\m}B_+ -
B_+{\pa}^{\m}A_+^*  \right) + \nonumber \\
&&
+ iq\biggl(A_+^*\wt{\c}_+ {\wt{\r}} +B_+\j_+\C {\l} \biggr) -  \left( qD+q^2
B_{\m} B^{\m}\right)A_+^*B_+ +\nonumber\\
&&
-F_-^{*}G_- - A_-^{*}\Box B_- - {1\over2} i {\j _-\C} {\sl{\pa}} \wt{\c}_- +
qB_{\m} \left({1\over2} i {\j_- \C}   \sm \wt{\c}_- + A_-^*{\pa}^{\m}B_- -
B_-{\pa}^{\m}A_-^*  \right) + \nonumber \\
&&
- iq\biggl(A_-^*\wt{\c}_- {\wt{\r}} +B_-\j_-\C {\l} \biggr) +  \left( qD- q^2
B_{\m} B^{\m}\right)A_-^*B_- +
\nonumber\\
&&\left.
+m \biggl(\frac12i\j_+\j_- - \frac12i\wt\c_+\wt\c_-
-A_+F_--A_-F_++B_+G_-+B_-G_+ \biggr)
\right\}+\mbox{h.c.}\;\;\;\;, \label{selfaction}
\eq
where $G_{\m\n}$ is the usual field-strength associated to $B_{\m}$.

Therefore, it can be easily seen, from (\ref{selfaction}), that the field
equation for $H^*_{\m\n}$ gives the self-duality of the field-strength
$G^{\m\n}$ :
\be
{\d S_{\sqed}^{\sd}\over \d H^*_{\m\n}}=0 \;\;\; \Longrightarrow \;\;\;
G^{\m\n}=\frac12 \epsilon^{\m\n\r\s}G_{\r\s}\;\;\;\;. \label{sdcond}
\ee

Since we are adopting the Wess-Zumino gauge, we can read directly
from the matter sector of (\ref{masssdqed}), the following set of local
$U(1)_{\a}$$\times$$U(1)_{\g}$ transformations {\cite{stau3qed,trab1}} :
\be
\d_{g} {A^*}_{\pm}={\pm}i q\b(x) {A^*}_{\pm} \;\;,\;\;\;
\d_{g} \j^{\C}_{\pm}={\pm}i q\b(x) \j^{\C}_{\pm} \aand
\d_{g} {F^*}_{\pm}={\pm}i q\b(x) {F^*}_{\pm} \;\;\;; \label{U(1)+-sym1}
\ee
\be
\d_{g} B_{\pm}={\mp}i q\b(x) B_{\pm} \;\;,\;\;\;
\d_{g} \wt\c_{\pm}={\mp}i q\b(x) \wt\c_{\pm} \aand
\d_{g} G_{\pm}={\mp}i q\b(x) G_{\pm} \;\;\;, \label{U(1)+-sym2}
\ee
where $\b$$\equiv$$\a$$-$$i\g$ is an arbitrary infinitesimal complex function.
The transformations for the gauge superfield components surviving the
Wess-Zumino gauge are as follows :
\be
\d_{g} \l=\d_{g} \wt\r=0\;\;,\;\;\;
\d_{g} D=0 \aand
\d_{g} B_{\m}= i\; \pa_{\m}\b \;\;\;.\label{gaugesupertrans}
\ee
Also, for the component fields of the multiplier superfield
(\ref{superspinor}), since $\d_{g}\Xi_{\a}$$=$$0$, we have :
\be
\d_{g} A_\a=\d_{g} F_\a=0\;\;,\;\;\;
\d_{g} E=0 \aand
\d_{g} H_{\m\n}= 0 \;\;\;.\label{multsupertrans}
\ee

Therefore, in the Wess-Zumino gauge, the $U(1)_{\g}$-symmetry is gauged by the
real part of $B_{\m}$ with real gauge function $\g$, whereas the
$U(1)_{\a}$-symmetry is gauged by its imaginary part with real gauge function
$\a$. By analysing the transformations (\ref{U(1)+-sym1}) and
(\ref{U(1)+-sym2}), a local Weyl-like symmetry $U(1)_{\g}$ naturally comes out
as one of the actual symmetries of the action (\ref{selfaction}). However, the
gauge field (the real part of $B_{\m}$) that gauges this symmetry will be
supressed in the process of dimensional reduction, then, such a symmetry, will
not persist in {\Ddd} {\cite{stau3qed}}.

Since ${\tau}_{3}$QED$_{1+2}$ coupled to a topological model in {\Ddd}
has been used in some theoreti\-cal approaches in Condensed Matter Physics
{\cite{domavro,qedtau3}} (and we are interested to obtain its $N$$=$$1$
supersymmetric version), it will be interesting to perform the dimensional
reduction proposed by Nishino {\cite{drscs}} on the action given by
eq.(\ref{selfaction}). Bearing in mind that this process should yield extended
supersymmetry {\cite{scherk,sohnius}}, some truncations will be needed in
order to remain with an $N$$=$$1$ supersymmetry in {\Ddd}, as well as to
suppress unphysical modes that will certainly appear after the dimensional
reduction are performed {\cite{stau3qed}}. These modes correspond to
negative-norm 1-particle states (ghosts) and they will be unavoidable in 3
dimensions, for the kinetic terms of the action (\ref{selfaction}) are totally
off-diagonal.

We perform the dimensional reduction{\footnote{After the dimensional reduction
is performed, the 3-dimensional metric becomes $\eta_{m n}$$=$$(+,-,-)$. Note
that, $\l$,
$\r$, $A$, $F$, $\j$ and $\c$ are now Dirac spinors in {\Ddd}.}} {\it{\`a la}}
Nishino {\cite{drscs}} from {\DDdd} to {\Ddd} on the action (\ref{selfaction}).
As a result, it can be found the following supersymmetric action in {\Ddd} :
\bq
S^{D=3}\!\!\!&=&\!\!\!\int{d^3\hat{x}}\left\{
{\frac \m2} \e^{klm}B^*_k G_{lm}+i{\frac \m2}{\ov A}{\gm {\pa}_m}\r-{\frac
\m2}{\ov F}\l+{\frac \m2} E^*D\;+\right.
\nonumber\\
&&
-\;F_+^{*}G_+ - A_+^{*}\Box B_+ - {1\over2} i {\ov\j _+} {\gm {\pa}_m} {\c}_+ -
 qB_{m} \left({1\over2} i {\ov\j_+ } \gm {\c}_+ + A_+^*{\pa}^{m}B_+ -
B_+{\pa}^{m}A_+^*  \right) + \nonumber \\
&&
 +\; q\biggl(A_+^* \ov{\c}_+\C {\r} -B_+\ov\j_+
{\l} \biggr) -  \left( qD+q^2B_{m} B^{m}\right)A_+^*B_+ +\nonumber \\
&&
-\;F_-^{*}G_- - A_-^{*}\Box B_- - {1\over2} i {\ov\j _-} {\gm {\pa}_m} {\c}_- +
 qB_{m} \left({1\over2} i {\ov\j_- } \gm {\c}_- + A_-^*{\pa}^{m}B_- -
B_-{\pa}^{m}A_-^*  \right) + \nonumber \\
&&
-\; q\biggl(A_-^* \ov{\c}_-\C {\r} -B_-\ov\j_-
{\l} \biggr) +  \left( qD-q^2B_{m} B^{m}\right)A_-^*B_- +\nonumber\\
&&\left.
-\;m \biggl(\frac12 \ov\j_+\C\j_- + \frac12\ov\c_+\C \c_-
+A_+F_-+A_-F_+-B_+G_--B_-G_+ \biggr)
\right\}+\mbox{h.c.} \;\;\;\;\;, \label{action3}
\eq
where the real parameter, $\m$, has dimension of mass. Notice that after the
dimensional reduction, the coupling constant $q$ has acquired dimension of
(mass)$^{1\over2}$.

Since the spectrum of the action given by eq.(\ref{action3}) will be spoiled by
the presence of negative-norm states, truncations will be needed in order to
suppress these unphysical modes. However, to identify the ghost fields to be
truncated, we must to diagonalize the whole free sector of the action
(\ref{action3}).

To perform the diagonalization of the free action (\ref{action3}), we need to
find some linear combinations of the fields. Therefore, by the same procedure
used for the case presented in Ref.{\cite{stau3qed}}, we have found the
following transformations :

\begin{enumerate}
\item{gauge sector :}
\be
A=\frac1{\sqrt{2}} \left(\x + \h \right) \aand \r=\frac1{\sqrt{2}}
\left(\x - \h \right) \;\;\;;
\ee
\be
F={\sqrt{2}} \left(\vf + \f \right) \aand \l={\sqrt{2}}
\left(\vf - \f \right) \;\;\;;
\ee
\be
E=\frac1{\sqrt{2}} \left(\wh E + \wh D \right) \aand D=\frac1{\sqrt{2}}
\left(\wh E - \wh D \right) \;\;\;;
\ee
\item{fermionic and bosonic matter sector :}
\be
\j_\pm=\frac1{\sqrt2}\left(\wh\j_\pm \mp {\wh\j_\mp}^{\rm c}+\wh\c_\pm \pm
{\wh\c_\mp}^{\rm c}\right) \aand \c_\pm=\frac1{\sqrt2}\left(\wh\c_\pm \pm
{\wh\c_\mp}^{\rm c}-\wh\j_\pm \pm {\wh\j_\mp}^{\rm c}\right) \;\;\;;
\ee
\be
A_\pm=\frac1{\sqrt2}\left[\frac1{\sqrt2}\left(\bv A_\pm \mp\bv A_\mp^*\right)
-\wh B_\pm\right] \aand
B_\pm=\frac1{\sqrt2}\left[\frac1{\sqrt2}\left(\bv A_\pm \mp\bv A_\mp^*\right)
+\wh B_\pm\right]\;\;\;;
\ee
\be
F_\pm=\frac1{\sqrt2}\left[\frac1{\sqrt2}\left(\bv F_\pm \mp\bv F_\mp^*\right)
+\wh G_\pm\right] \aand
G_\pm=-\frac1{\sqrt2}\left[\frac1{\sqrt2}\left(\bv F_\pm \mp\bv F_\mp^*\right)
-\wh G_\pm\right]\;\;\;.
\ee
\end{enumerate}

By replacing these field redefinitions into the action (\ref{action3}), one
ends up with a diagonalized action, where the fields, $\h$, $\wh\c_+$,
$\wh\c_-$, $\wh B_+$ and $\wh B_-$ appear like ghosts in the framework of an
$N$$=$$2$-supersymmetric model. Therefore, in order to suppress these
unphysical modes, truncations must be performed. Bearing in mind that we are
looking for an $N$$=$$1$ supersymmetric 3-dimensional model (in the Wess-Zumino
gauge), truncations have to be imposed on the ghost fields, $\h$, $\wh\c_+$,
$\wh\c_-$, $\wh B_+$ and $\wh B_-$. To keep $N$$=$$1$ supersymmetry
in the Wess-Zumino gauge, we must simultaneously truncate the component fields,
$\wh G_+$, $\wh G_-$, $\wh D$, $\wh E$, $\x$, $\f$, $a_m$ and $\t$
{\footnote{The $a_m$ field is the
real part of $B_m$, since we are assuming $B_m$$=$$a_m$$+$$iA_m$. Also, as
$\vf$ is a Dirac spinor, it can be written in terms of two Majorana spinors
in the following manner: $\vf$$=$$\t$$-$$i\wh\l$.}} . Now, the choice of
truncating $a_m$,
instead of $A_m$, is based on the analysis of the couplings to the matter
sector: $A_m$ couples to both scalar and fermionic matter and we interpret it
as the photon field in 3 dimensions.

After performing these truncations, and omitting the $(\widehat{\;\;\;})$ and
$(\bv{\;\;\;})$ symbols, we find the following action in {\Ddd} :
\bq
S_{\t_3{\rm QED}}^{\rm SCS}\!\!\!\!&=&\!\!\!\!\int{d^3\hat{x}}\left\{
{\m} \e^{klm}A_k F_{lm}- 2\m{\ov \l}\l \;+\right.
\nonumber\\
&&
-\;A_+^{*}\Box A_+ - A_-^{*}\Box A_- + i {\ov\j _+} {\gm {\pa}_m} {\j}_+ + i
{\ov\j _-} {\gm {\pa}_m} {\j}_- + F_+^{*}F_+ + F_-^{*}F_- + \nonumber \\
&& -\;q A_{m}\biggl({\ov\j_+ } \gm {\j}_+ - {\ov\j_- } \gm {\j}_- +
iA_+^*{\pa}^{m}A_+ - iA_-^*{\pa}^{m}A_- - iA_+{\pa}^{m}A_+^* +
iA_-{\pa}^{m}A_-^* \biggr) + \nonumber \\
&&
 -\;iq \biggl(A_+ \ov{\j}_+ {\l} - A_-\ov\j_- {\l} - A_+^{*} \ov{\l} {\j}_+ +
A_-^{*} \ov{\l} {\j}_- \biggr) +  q^2 A_{m} A^{m}\left(A_+^*A_+ + A_-^*A_-
\right) +\nonumber \\
&&\left.
-\;m \biggl( \ov\j_+\j_+ - \ov\j_- \j_- + A_+^{*}F_+ - A_-^{*}F_- + A_+F_+^{*}
- A_-F_-^{*} \biggr)
\right\} \;\;\;\;\;\;\;\;\;\;; \label{action3diag}
\eq
hence, we conclude that this is a supersymmetric extension of a
parity-preserving action minimally coupled to a Chern-Simons field
{\cite{djt,domavro,qedtau3}}. However, to
render our claim more explicit, we are going next to rewrite
(\ref{action3diag}) in terms of the superfields of $N$$=$$1$ supersymmetry in 3
dimensions.

In order to formulate the $N$$=$$1$ super-Chern-Simons coupled to the
${\tau}_{3}$QED (\ref{action3diag}) in terms of superfields, we refer to the
work by Salam and
Strathdee {\cite{Salam}}. Extending their ideas to our case in {\Ddd}, the
elements of superspace are labeled by $(x^m,\q)$, where $x^m$ are the
space-time coordinates and the fermionic coordinates, $\q$, are Majorana
spinors, $\q\C$$=$$\q$. {\footnote{The adjoint and charge-conjugated spinors
are defined by $\ol\j$$=$$\j^\dg \g^0$ and $\j\C$$=$$C\ol\j^T$, repectively,
where $C$$=$$-\sy$. The $\g$-matrices we are using arised from the dimensional
reduction to {\Ddd} are: $\g^m$$=$$(\sx,i\sy,-i\sz)$. Note that for any
spinorial objects, $\j$ and $\c$, the product $\ol\j\c$ denotes $\ol\j_a
\c_a$.}}

Now, we define the $N$$=$$1$ complex scalar
superfields with opposite $U(1)$-charges, $\F_\pm$, as
\be
\F_\pm=A_\pm+\ol\q\j_\pm-\frac12\ol\q\q F_\pm \label{scalar3a} \aand
\F_\pm^\dg=A_\pm^*+\ol\j_\pm\q-\frac12\ol\q\q F_\pm^* \;\;\;\;,
\label{scalar3b}
\ee
where $A_+$ and $A_-$ are complex scalars, $\j_+$ and $\j_-$ are Dirac spinors
and $F_+$ and $F_-$ are complex auxiliary scalars.
Their gauge-covariant derivatives read :
\be
\nabla_a\F_\pm=\left(D_a\mp iq\G_a\right)\F_\pm
\aand\ol\nabla_a\F^\dg_\pm=\left(\ol D_a\pm iq\ol\G_a\right)\F^\dg_\pm
\;\;\;\;, \label{deriv3}
\ee
where $D_a$$\equiv$${\ol\pa}_a$$-$$i(\g^m\q)_a\pa_m$ and ${\ol
D}_a$$\equiv$$\pa_a$$+$$i(\ol\q\g^m)_a\pa_m$ . The gauge superconnection,
$\G_a$, is written in the Wess-Zumino gauge as
\be
\G_a=i(\g^m\q)_aA_m+\ol\q\q\l_a \label{gauge3a} \aand
\ol{\G}_a=-i(\ol\q\g^m)_aA_m+\ol\q\q\ol\l_a \;\;\;\;, \label{gauge3b}
\ee
with field-strength superfield, $W_a$, given by
\be
W_a=\frac12{\ol D}_b D_a\G_b \;\;\;\;. \label{strength3}
\ee

By using the previous definitions of the superfields, (\ref{scalar3a}),
(\ref{gauge3a}) and (\ref{strength3}), and the gauge-covariant derivatives,
(\ref{deriv3}), we found how to build up the $N$$=$$1$ super-${\tau}_{3}$QED
action coupled to a super-Chern-Simons term, in superspace; it reads :
\be
S_{\t_3{\rm QED}}^{\rm SCS}\!=\!\int\! d\hat v\left\{{2\m}
({\ol \G}W)-(\ol\nabla\F_+^\dg)(\nabla\F_+)-(\ol\nabla\F_-^\dg)
(\nabla\F_-)+2m(\F_+^\dg\F_+-\F_-^\dg\F_-)\right\} \,,
\label{superqedtau3}
\ee
where the superspace measure we are adopted is $d\hat v$$\equiv$$d^3\hat x
d^2\q$ and the Berezin integral is taken as $\int\!d^2\q
$$=$$\frac14\ol\pa\pa$.

Our final conclusion is that the massive Abelian $N$$=$$1$ super-{\QED} coupled
to a self-dual supermultiplet
as proposed in ref.{\cite{trab1}}, shows interesting features when an
appropriate
dimensional reduction is performed. The dimensional reduction {\it{\`a la}}
Nishino we have applied to our problem becomes very attractive, since, after
doing some truncations to avoid non-physical modes, $N$$=$$1$
super-Chern-Simons coupled to a parity-preserving matter sector
(super-${\tau}_{3}$QED) is obtained as a final result.

%%%%%%%%%%%%%%%%%%%%%%%%%%%%%%%%%%
\subsection*{Acknowledgements}
%%%%%%%%%%%%%%%%%%%%%%%%%%%%%%%%%
Dr. J.A. Helay\"el-Neto and Dr. O. Piguet are kindly acknowledged for
suggestions and patient discussions.  O.M.D.C. and M.A.D.A. express their
gratitude to the Organizing Committee of the {\it Spring School and Workshop on
String Theory, Gauge Theory and Quantum Gravity '95} for the kind hospitality
at the {\it International Centre for Theoretical Physics (ICTP)}. L.P.C. is
grateful to Prof. M.A. Virasoro for the kind hospitality at {\it ICTP}. {\it
CNPq-Brazil} is acknowledged for the invaluable financial help.

\end{document}